\documentclass{article}
\usepackage{spconf,amsmath,graphicx,multirow}
\usepackage{url}

\title{Lossless Compression in HEVC with Integer-to-Integer Transforms}
\name{Fatih Kamisli}
\address{Middle East Technical University\\
Ankara, Turkey}
\begin{document}
%
\maketitle
\begin{abstract}
Many approaches have been proposed to support lossless coding within video coding standards that are primarily designed for lossy coding. The simplest approach is to just skip transform and quantization and directly entropy code the prediction residual, which is used in HEVC version 1. However, this simple approach is inefficient for compression. More efficient approaches include processing the residual with DPCM prior to entropy coding. This paper explores an alternative approach based on processing the residual with integer-to-integer (i2i) transforms. I2i transforms map integers to integers, however, unlike the integer transforms used in HEVC for lossy coding, they do not increase the dynamic range at the output and can be used in lossless coding. Experiments with the HEVC reference software show competitive results.
\end{abstract}

\section{Introduction}
\label{sec:intro}

Image\footnote{This research was supported by Grant 113E516 of T\"{u}bitak.} and video compression can be performed lossless or lossy. 
In lossless compression, the reconstructed image or video data is identical to the original visual data, and in lossy compression, some amount of degradation in the reconstruction is allowed to achieve higher compression. The emerging video coding standard HEVC \cite{HEVC} or the widely deployed standard H.264/AVC \cite{Luthra264} support both lossy and lossless compression. 

Lossy compression in these standards is achieved with a block-based approach. First, a block of pixels are predicted using pixels from a previously coded frame (inter prediction) or using pixels from previously coded regions of the current frame (intra prediction). The prediction is in many cases not accurate and as the next step, the block of prediction errors are computed and transformed to remove any remaining spatial redundancy. Finally, the transform coefficients are quantized and entropy coded together with other relevant side information such as prediction modes. 

It is desirable to support lossless compression using this lossy coding architecture with as little modification as possible so that encoders/decoders can also support lossless compression without any significant complexity increase. The simplest approach is to just skip the transform and quantization steps, and directly entropy code the block of prediction errors. This approach is indeed used in HEVC version 1 \cite{HEVC}. While this is a simple and low-complexity approach, it is well known that in many regions of images/videos, residuals of block-based prediction are not sufficiently decorrelated and directly entropy coding these prediction errors is inefficient for compression. Hence, a large number of approaches have been proposed to improve the compression performance.

The majority of the proposed approaches can be grouped into two groups. Approaches in the first group perform the standard block-based prediction and then apply differential pulse code modulation (DPCM) on the prediction error block to further decorrelate it. The output is then fed to the entropy coder. The DPCM is a pixel-by-pixel prediction algorithm and many variations of it have been proposed \cite{sulivanDPCM,cross,Secondary}. In the second group of approaches, which are applied to only lossless intra coding, the block-based intra prediction step is replaced directly with a pixel-by-pixel prediction approach \cite{SAP,DC2,Templatebased}. These approaches are discussed in more detail in the next section.

This paper explores an alternative approach for lossless compression within HEVC. In this approach, the residuals of block-based prediction are processed with integer-to-integer (i2i) transforms. Integer-to-integer transforms map integers to integers. However, unlike the integer transforms used in HEVC for lossy coding, they do not increase the dynamic range at the output and can therefore be easily employed in lossless coding. This paper uses a computationally efficient i2i approximation of the 4-point DCT to process both intra and inter prediction residual blocks. While there are many papers that employ i2i transforms in lossless or lossless-to-lossy image compression \cite{i2iLU,binDCT,L2L}, we could not come across a work which explores i2i transforms for lossless compression of prediction residuals within video coding standards H.264/AVC or HEVC. 

The remainder of the paper is organized as follows. In Section \ref{sec:pr}, a brief overview of related previous research on lossless video compression is provided. Section \ref{sec:i2i} discusses i2i transforms and Section \ref{sec:exp} presents experimental results of using an i2i transform within HEVC for lossless compression. Finally, Section \ref{sec:con} concludes the paper.

\section{Review of Related Previous Research On Lossless Video Compression}
\label{sec:pr}



\subsection{Methods based on residual DPCM}
\label{ssec:rdpcm}
Methods based on residual differential pulse code modulation (RDPCM) first perform the default block-based prediction and then process the prediction error block further with a DPCM method, i.e. a pixel-by-pixel prediction method. The literature contains many variations of this general RDPCM approach \cite{sulivanDPCM,cross,Secondary}.

One of the earliest of such methods was proposed in \cite{sulivanDPCM} for lossless intra coding in H.264/AVC. Here, first the block-based spatial prediction is performed, and then a simple pixel-by-pixel differencing operation is applied on the residual pixels in only horizontal and vertical intra prediction modes. In horizontal mode, from each residual pixel, its left neighbor is subtracted and the result is the RDPCM pixel of the block. Similar differencing is performed along the vertical direction in the vertical intra mode. Note that the residuals of other angular modes are not processed in \cite{sulivanDPCM}. 

The same RDPCM method as in \cite{sulivanDPCM} is now included in HEVC version 2 \cite{HEVCv2,L0117} for intra and inter coding. In inter coding, RDPCM is applied either along the horizontal or vertical direction or not at all, and a flag is coded in each transform unit (TU) to indicate if it is applied, and if so, another flag is coded to indicate the direction. In intra coding, RDPCM is applied only when prediction mode is either horizontal or vertical and no flag is coded since the RDPCM direction is inferred from the intra prediction mode.


\subsection{Methods based on pixel-by-pixel prediction}


Methods based on pixel-by-pixel prediction are applied to only intra coding in lossless compression. Since the transform is skipped in lossless coding, a pixel-by-pixel intra prediction approach can now be used instead of a block-based approach for more efficient prediction. The literature contains many such methods \cite{SAP,DC2,Templatebased}. While these methods can provide the best compression performance \cite{3tap}, their distinguishing property can also be a drawback : their pixel-based nature is not congruent with the block-based architecture of video codecs and introduces undesired pixel-based dependencies in the prediction architecture \cite{SAP}.


\subsection{Methods based on modified entropy coding}
\label{ssec:entropy}
In lossy coding, transform coefficients of prediction residuals are entropy coded, while in lossless coding, the prediction residuals are entropy coded. Considering the difference of the statistics of quantized transform coefficients and prediction residuals, several modifications in entropy coding were proposed for lossless coding \cite{kim2010efficient,CABAC3,CABAC1}. The HEVC version 2 includes reversing the scan order of coefficients, using a dedicated context model for the significance map and other tools \cite{HEVCv2,N0044}.

\section{Integer-to-integer (i2i) transforms}
\label{sec:i2i}

Integer-to-integer (i2i) transforms map integer inputs to integer outputs and are invertible. However, unlike the integer transforms in HEVC \cite{HEVCtr}, which also map integers to integers, they do not increase the dynamic range at the output. Therefore they can be easily used in lossless compression. 

A significant amount of work on i2i transforms is done to develop i2i approximations of the discrete cosine transform (DCT). There are a number of ways to develop i2i DCTs and in all of them the common approach is to decompose the DCT into elementary operations \cite{binDCT,i2iSERM,i2iLU}. Each elementary operation is then modified to map integers to integers through rounding operations. 
Perhaps the most popular method, due its to lower computational complexity, is to use the factorization of the DCT into plane rotations and butterfly structures \cite{Chen,Loeffler,binDCT}. This is also the method that is used in this paper.

\subsection{I2i DCT through factorization of DCT into plane rotations and butterflies and the lifting scheme}
\begin{figure}[b]
  \begin{center}
    \includegraphics[trim=40 600 300 50,clip,width=0.9\linewidth]{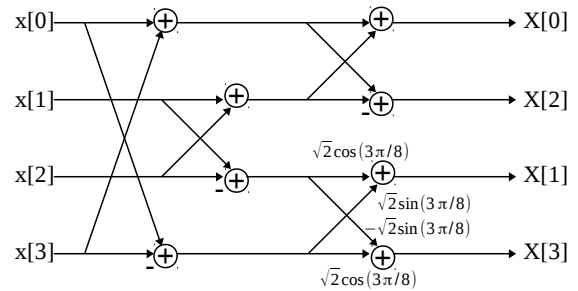}
    \caption{Factorization of 4-point DCT.}
    \label{fig:LoeffFact4}
  \end{center}
\end{figure}

Two well-known factorizations of the DCT into plane rotations and butterflies are the Chen's and Loeffler's factorizations \cite{Chen,Loeffler}. Loeffler's 4-point DCT factorization is shown in Figure \ref{fig:LoeffFact4}. It contains three butterflies and one plane rotation. Note that the output samples in Figure \ref{fig:LoeffFact4} need to be scaled by $1/2$ to obtain an orthogonal DCT.

The butterfly structure shown in Figure \ref{fig:LoeffFact4} maps integers to integers because the output samples are the sum and difference of the inputs. It is also easily invertible by itself and dividing the output samples by 2. 

\begin{figure}
\begin{center}
\begin{minipage}{0.49\linewidth}
\centering
\includegraphics[trim=65 680 420 55,clip,width=\linewidth]{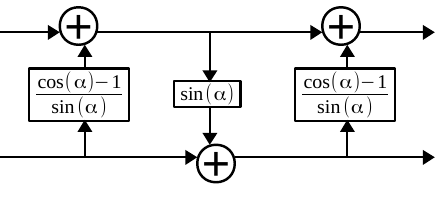}
\centerline{{\small (a) Forward Rotation}}
\end{minipage}
\begin{minipage}{0.49\linewidth}
\centering
\includegraphics[trim=65 680 420 55,clip,width=\linewidth]{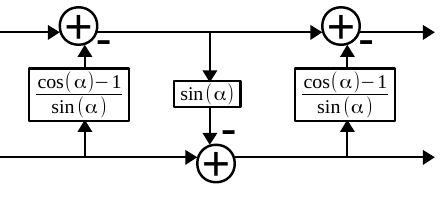}
\centerline{{\small (b) Inverse Rotation}}
\end{minipage}
\end{center}
\caption{Decomposition of plane rotation into three lifting steps. (a) Forward Rotation (b) Inverse rotation.}
\label{fig:rotlif}
\end{figure}

The plane rotation in Figure \ref{fig:LoeffFact4} can be modified as follows so that it also maps integers to integers. A plane rotation can be represented with the 2x2 matrix below and it can be decomposed into three lifting steps as follows :
\begin{equation}
\left[   \begin{array}{ c c }
     \cos(\alpha) & -\sin(\alpha) \\
     \sin(\alpha) &  \cos(\alpha) 
  \end{array} \right]
= 
\left[   \begin{array}{ c c }
     1 & p \\
     0 & 1 
  \end{array} \right]
\left[   \begin{array}{ c c }
     1 & 0 \\
     u & 1 
  \end{array} \right]  
\left[   \begin{array}{ c c }
     1 & p \\
     0 & 1 
  \end{array} \right].  
\label{eq:rot}  
\end{equation}
Here, $p=\dfrac{\cos(\alpha)-1}{\sin(\alpha)}$ and $u=\sin(\alpha)$. A graphical representation of the decomposition is shown in Figure \ref{fig:rotlif}-a.

Each lifting step can be inverted with another lifting step:
\begin{equation}
\left[   \begin{array}{ c c }
     1 & p \\
     0 & 1 
  \end{array} \right]^{-1}
=
\left[   \begin{array}{ c c }
     1 & -p \\
     0 & 1 
  \end{array} \right]  
,
\left[   \begin{array}{ c c }
     1 & 0 \\
     u & 1 
  \end{array} \right]^{-1}
=
\left[   \begin{array}{ c c }
     1 & 0 \\
     -u & 1 
  \end{array} \right].   
\label{eq:lif}  
\end{equation}
In other words, each lifting step is inverted by subtracting out what was added in the forward lifting step (see Figure \ref{fig:rotlif}-b). Notice that each lifting step is still invertible even if the multiplication of the input samples with floating point $p$ or $u$ are rounded to integers, as long as the same rounding operation is applied in both forward and inverse lifting steps. This implies that each lifting step can map integers to integers and is easily invertible.

Notice that each lifting step in the above factorization requires floating point multiplications since $p$ and $u$ are in general not integers. To avoid floating point multiplications, the lifting factors $p$ and $u$ can be approximated with with rationals of the form $k/2^m$ ($k$ and $m$ are integers), which can be implemented with only addition and bitshift operations.
 
\begin{figure}
  \begin{center}
    \includegraphics[trim=65 680 420 55,clip,width=0.5\linewidth]{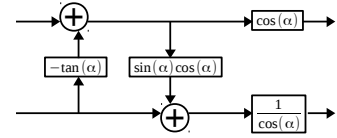}
    \caption{Decomposition of plane rotation into two lifting structures and two scaling factors.}
    \label{fig:rotlif2}
  \end{center}
\end{figure}

\begin{figure}
  \begin{center}
    \includegraphics[trim=40 600 300 50,clip,width=0.9\linewidth]{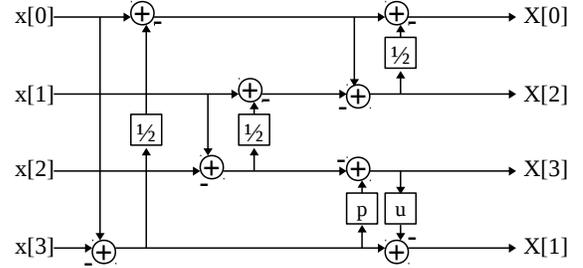}
    \caption{I2i DCT with minimized dynamic range for lossless compression.}
    \label{fig:binDCT4}
  \end{center}
\end{figure}

A plane rotation can be represented by three lifting steps as in Figure \ref{fig:rotlif} or by two lifting steps and two scaling factors as shown in Figure \ref{fig:rotlif2} \cite{binDCT}. Using two lifting steps per plane rotation reduces the complexity. The two scaling factors can be combined with other scaling factors at the output, creating a scaled i2i DCT. The scaling factors at the output can be absorbed into the quantization stage in lossy coding. In lossless coding, all scaling factors can be omitted. However, care is needed when omitting scaling factors since for some output samples, the dynamic range may become too high when scaling factors are omitted. For example, in Figure \ref{fig:LoeffFact4}, the DC output sample becomes the sum off all input samples when scaling factors are omitted, however, it may be preferable that it is the average of all input samples. This can improve the entropy coding performance. Hence it is common in lossless coding to replace butterflies of Figure \ref{fig:LoeffFact4} with lifting steps as shown in Figure \ref{fig:binDCT4} \cite{binDCT,L2L}.

The $p$ and $u$ scaling factors in Figure \ref{fig:binDCT4} can be chosen as rationals of the form $k/2^m$ ($k$ and $m$ are integers), to be implemented with only addition and bitshift operations. Note that bitshift operation implicitly includes a rounding operation, which is necessary for mapping integers to integers, as discussed above. $k$ and $m$ can be chosen depending on the desired accuracy to approximate the DCT and the desired level of computational complexity. 

\subsection{I2i DCT within HEVC}
\label{ssec:i2ihevc}
This paper uses the i2i DCT approximation shown in Figure \ref{fig:binDCT4} to explore i2i transforms in lossless coding within HEVC. The scaling factors $p$ and $u$ are chosen as $p=1/2$ and $u=1/2$. One reason for this choice is that it allows implementing the i2i DCT with only addition and bitshift operations. Another reason is that, since the i2i transform is used on prediction residuals, approximating the DCT very well is not necessary since prediction residuals have typically smaller correlation than image pixels.

The i2i DCT approximation in Figure \ref{fig:binDCT4} is used along first the horizontal and then the vertical direction to obtain an i2i 2D DCT. The i2i 2D DCT is used in lossless compression to transform both intra and inter prediction residuals of luma and chroma pictures in only 4x4 transform units (TU). The transform coefficients are directly fed to the entropy coder without quantization. In larger TUs, the default HEVC processing is used. Notice that in lossless coding, the encoder choses 4x4 TUs much more frequently (even at large resolutions) than other TUs. 
Exploring i2i transforms in larger TUs is part of our plans for future work.

\section{Experimental Results}
\label{sec:exp}

The explored i2i transform approach is implemented into the HEVC version 2 Range Extensions (RExt) reference software (HM-15.0+RExt-8.1) \cite{HMref} to provide experimental results for the i2i transform approach. The following systems are derived from the reference software and compared in terms of lossless compression performance and complexity :
\begin{itemize}
  \itemsep0em
  \item HEVCv1
  \item HEVCv2
  \item I2I
  \item I2I+RDPCM.
\end{itemize}

The HEVCv1 system represents HEVC version 1, which just skips transform and quantization for lossless coding, as discussed in Section \ref{sec:intro}. The HEVCv2 system represents HEVC version 2, and includes all available RExt tools, such as RDPCM, reversing the scan order, a dedicated context model for the significance map and other tools \cite{HEVCv2,N0044} as discussed in Sections \ref{ssec:rdpcm} and \ref{ssec:entropy}.

The remaining two systems, I2I and I2I+RDPCM employ the i2i transform discussed in Section \ref{ssec:i2ihevc}. The i2i transform is used in intra and inter coded blocks in only 4x4 transform units (TU). In larger TUs, the default processing of HEVC version 2 is used. 

In the I2I system, the RDPCM system of the reference software is disabled in 4x4 TUs, and is replaced with the 4x4 i2i transform. In intra coding, the i2i transform is applied to the residual TU of all intra prediction modes. In inter coding, the i2i transform is either applied or not applied on the prediction residual TU, and is indicated to the decoder with a flag, similar to HEVC version 2, as discussed in Section \ref{ssec:rdpcm}.

In the I2I+RDPCM system, the i2i transform and RDPCM methods are combined in intra coding. In other words, in intra coding of 4x4 TUs, the RDPCM method of HEVC version 2 is used if the intra prediction mode is equal to horizontal or vertical mode, and the i2i transform method is used for other intra prediction modes. Inter coding remains the same as in the I2I system. Table \ref{tb:systems} summarizes the processing in all systems.

\begin{table}[tbh]
\centering
\caption{Processing of 4x4 block prediction residuals prior to entropy coding in each system.}
\label{tb:systems}
\begin{tabular}{l|ccccc}
\hline
                      & HEVCv1  & HEVCv2  & I2I     & I2I+RDPCM  \\ \hline \hline
hor/ver intra  &   -     & h/v rdpcm   & i2i     & h/v rdpcm       \\ \hline
other intra    &   -     & -    & i2i     & i2i       \\ \hline
\multirow{3}{*}{inter}    &     & h rdpcm,   & i2i,     & i2i,     \\
                          &   - & v rdpcm,   & -        & -       \\
                          &     & -             &          &         \\ \hline
\end{tabular}
\end{table}

For the following experimental results, the common test conditions in \cite{commontest} are followed, except that only the first 32 frames are coded due to our limited computational resources. All results are shown in Table \ref{tb:res}, which includes average percentage ($\%$) bitrate reduction and encoding/decoding time of HEVCv2, I2I, and I2I+RDPCM systems with respect to HEVCv1 system for All-Intra-Main, Low-Delay-Main and Random-Access-Main settings. 

\begin{table*}[tbh]
\centering
\caption{Average percentage ($\%$) bitrate reduction and encoding/decoding time of HEVCv2, I2I, and I2I+RDPCM systems with respect to HEVCv1 in lossless coding for All-Intra-Main, Low-Delay-Main and Random-Access-Main settings.}
\label{tb:res}
\begin{tabular}{l|ccc|ccc|ccc}
\hline
      & \multicolumn{3}{|c|}{All-Intra-Main} & \multicolumn{3}{|c|}{Low-Delay-Main} & \multicolumn{3}{|c}{Random-Access-Main} \\ \hline \hline 
      & HEVCv2  & I2I  & I2I+RDPCM           & HEVCv2  & I2I  & I2I+RDPCM            & HEVCv2  & I2I  & I2I+RDPCM \\ \hline \hline
Class A   & 7.2 & 10.2 & 11.2                & 4.4     & 6.0  & 6.3                  &  4.4    & 6.0  &   6.3     \\ \hline
Class B   & 4.7 &  4.3 &  5.1                & 2.0     & 1.7  & 1.9                  &  2.0    & 1.8  &   2.0     \\ \hline
Class C   & 5.4 &  3.8 &  5.1                & 2.4     & 2.0  & 2.4                  &  2.5    & 2.2  &   2.6     \\ \hline
Class D   & 7.6 &  6.5 &  8.2                & 2.3     & 2.2  & 2.4                  &  2.6    & 2.4  &   2.6     \\ \hline
Class E   & 8.2 &  8.3 &  9.7                & 4.3     & 3.5  & 4.0                  &  4.2    & 3.3  &   3.8     \\ \hline
Average   & 6.4 &  6.4 &  7.6                & 3.0     & 3.0  & 3.3                  &  3.0    & 3.0  &   3.4     \\ \hline
Enc. Time & $94\%$ & $96\%$ & $97\%$         & $95\%$ & $94\%$ & $94\%$              & $97\%$  & $95\%$ & $96\%$ \\ \hline
Dec. Time & $94\%$ & $94\%$ & $97\%$         & $96\%$ & $95\%$ & $98\%$              & $97\%$  & $95\%$ & $99\%$ \\ \hline
\end{tabular}
\end{table*}

Consider first the results for All-Intra-Main coding settings. HEVCv2, I2I, and I2I+RDPCM systems achieve $6.4$, $6.4$ and $7.6$ percent overall average bitrate reduction over HEVCv1, respectively. For Class A sequences, which include sequences with the largest resolution (2560x1600), systems employing i2i transforms (I2I and I2I+RDPCM systems) achieve significantly larger bitrate reductions than the HEVCv2 system. For the other classes, the I2I+RDPCM system is typically slightly better than the HEVCv2 system, which in turn is typically slightly better than the I2I system.

We note that the results for All-Intra-Main coding settings may look counter-intuitive at first, since the i2i transforms are applied at only the smallest 4x4 TUs but the largest coding gains by I2I or I2I+RDPCM systems are achieved in Class A sequences, which have the largest resolution sequences. However, while the tendency of the encoder to choose smaller block-sizes (i.e. 4x4 TUs) less often at higher resolutions than in lower resolutions is significant for lossy coding (due to RD-optimized mode decisions), it becomes less significant at high bitrates or lossless coding. This is due to following reasons. First, since there is no quantization in lossless coding, RDPCM or i2i transforms can not obtain many zero-coefficients like in lossy coding with quantization, and thus prediction becomes very important in lossless coding, and prediction is most effective at the smallest available block-size. Furthermore, any overhead bitrate due to using smallest block-size PUs or TUs becomes negligible in lossless coding compared to the bitrate of coding the actual residual. In summary, even at higher resolutions, the encoder chooses 4x4 blocks for prediction or transformation very often in lossless coding.

The actual reason behind the better coding gains with i2i transforms at higher resolutions can be explained by the statistical characteristics of the residuals. In larger resolutions, a similar sized image or residual block (e.g. 4x4 block) is more likely to have higher spatial correlation than in smaller resolutions. For a stronger correlated signal, a transform is expected to provide larger coding gains than a DPCM technique. For less correlated signals, the coding gain difference is expected to diminish, and if the used transform is an i2i one, its coding gain can drop below that of DPCM. 

RDPCM is a very effective method for horizontal and vertical intra modes, because for these modes, block-based prediction followed by RDPCM becomes overall equivalent to pixel-by-pixel prediction, which is known to be a very effective method \cite{SAP,3tap}. However, in HEVC, RDPCM is used  to process only residuals of horizontal and vertical intra modes, while the residuals of other intra modes are not processed at all. Hence, as can be seen from the All-Intra-Main results in Table \ref{tb:res}, depending on the image/video characteristics, the i2i transform applied to all intra modes (I2I system) can obtain larger coding gains, or the RDPCM method  applied to only horizontal and vertical modes (HEVCv2 system) can obtain larger coding gains. However, when i2i transform and RDPCM methods are combined (I2I+RDPCM system), then the best coding gain is achieved almost exclusively as shown in Table \ref{tb:res}.


Consider next the results for Low-Delay-Main and Random-Access-Main coding settings. In both of these settings, HEVCv2, I2I, and I2I+RDPCM systems achieve $3.0$, $3.0$ and $3.3$ percent overall average bitrate reduction over HEVCv1, respectively. Again, for Class A sequences, systems employing i2i transforms (I2I and I2I+RDPCM systems) achieve significantly larger bitrate reductions than the HEVCv2 system. For the other classes, all three systems achieve bitrate reductions that are typically close to each other. Notice that the residuals of inter prediction are typically much less correlated than those of intra prediction, and thus the coding gains achievable by RDPCM or i2i transform methods are much lower for inter coding than those of intra coding.

The average encoding and decoding times are also shown in Table \ref{tb:res}. They are compared to those of HEVCv1 in percentages. The HEVCv2, I2I, and I2I+RDPCM systems achieve up to $6\%$ lower encoding or decoding times than HEVCv1, despite their additional processing of the residuals, mainly due to their lower bitrates which allow the complex entropy coding/decoding to finish faster.

\section{Conclusions}
\label{sec:con}
This paper explored an alternative approach for lossless video coding based on integer-to-integer (i2i) transforms within HEVC. I2i transforms map integers to integers without increasing the dynamic range at the output and were used in this paper to transform intra and inter prediction residuals of luma and chroma pictures in only 4x4 transform units (TU). Experimental results showed competitive performance with respect to other major methods, such as RDPCM, in terms of both compression performance and complexity. Several directions for future research are possible, such as exploring use of i2i transforms also in larger TUs and adaptively turning on/off the i2i transforms in each TU.

%


\end{document}